\documentclass[twocolumn,english,aps,prb,floatfix,preprintnumbers,showpacs,amsfonts,amssymb,superscriptaddress]{revtex4}
\usepackage{ae}
\usepackage{aecompl}
\usepackage[T1]{fontenc}
\usepackage[latin1]{inputenc}
\usepackage{amsmath}
\usepackage{amssymb}

\makeatletter
\usepackage{graphicx}
\usepackage{amscd}
\usepackage{bm}

\usepackage{babel}
\makeatother
\begin{document}

\title{Absence of a true long-range orbital order in a two-leg Kondo ladder }

\author{J.~C.~Xavier}

\affiliation{Instituto de F\'{\i}sica, Universidade Federal de Uberl\^{a}ndia,
Caixa Postal 593, 38400-902 Uberlândia, MG, Brazil}

\author{A.~L. Malvezzi}

\affiliation{Departamento de F\'{\i}sica, Faculdade de Ci\^{e}ncias, Universidade
Estadual Paulista, Caixa Postal 473, 17015-970 Bauru, SP, Brazil}

\date{\today{}}

\begin{abstract}
We investigate, through the density-matrix renormalization group and
the Lanczos technique, the possibility of a two-leg Kondo ladder present
an incommensurate orbital order. Our results indicate a staggered
short-range orbital order at half-filling. Away from half-filling
our data are consistent with an incommensurate quasi-long-range orbital
order. We also observed that an interaction between the localized
spins enhances the rung-rung current correlations. 
\end{abstract}

\pacs{71.10.Pm, 75.10.-b, 75.30.Mb}

\maketitle

\section{INTRODUCTION}

In 1985, it was observed that the heavy fermion superconductor $URu_{2}Si_{2}$
presents a second order phase transition at 17.5K.\cite{Palstraetal}
This phase transition is characterized by sharp features in the specific
heat \cite{Palstraetal} and several others thermodynamic properties
(see, \emph{e.g.}, Ref. \onlinecite{colemannatu} and References therein).
The large entropy loss associated in this phase transition is equivalent
 to an ordered moment of about 0.5$\mu_{B}.$
However, the size of staggered moment measured by neutrons scattering
measurements is $m\sim0.03\mu_{B}$.\cite{Palstraetal} The order
parameter associated with this phase transition is, at the present
moment, not established and it is challenging to discover the nature
of the hidden order behind the transition.

Many theoretical groups have proposed several kinds of hidden order.
\cite{helicityorder,colemannatu,agterbeg,hiroaki,santini,fazekasoctu}
But, until now, experiments were not able to establish which is the
correct one. Certainly, also from the theoretical point of view, more
studies are needed to clarify the correct order associated with this
mysterious phase transition. In this front, we present here a numerical
study of a microscopic model for the heavy fermion systems. 

In this work we focus on the order parameter proposed a few years ago
by Chandra and collaborators. \cite{colemannatu} They suggested the
existence of a hidden incommensurate orbital order in the heavy fermion
$URu_{2}Si_{2}$ below the second order phase transition. The orbital
order phase is associated with currents circulating around the plaquettes,
as illustrated in Fig. 1. In the case of $URu_{2}Si_{2}$, this currents
produce a very week orbital moment $~0.02\mu_{B}$ that explains the
large entropy loss.\cite{colemannatu}

\begin{figure}
\begin{center}\includegraphics[scale=0.42]{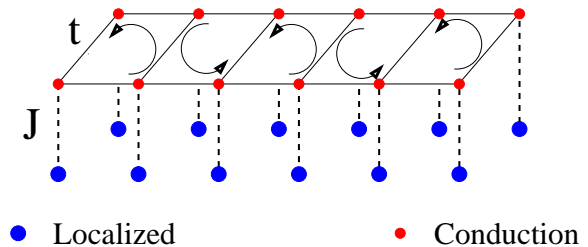}\end{center}

\caption{\label{fig1} (Color online) A schematic representation of the two-leg
Kondo ladder. It also shows the circulating currents around the plaquettes
(in this example a staggered one).}
\end{figure}

Very recently, neutron scattering measurements were unable to detect
the orbital order in the heavy fermion $URu_{2}Si_{2}$.\cite{orbitalexp}
Although the orbital order was not detected it is not possible yet
to discard it as the hidden order due to the resolution limitation
of the experiments performed. Note that the orbital order is expected
to be $~50$ times smaller than the spin order.\cite{colehiden}

Our goal in this work is to investigate the existence of an incommensurate
orbital order in the Kondo Lattice model (KLM). This model is the
simplest one believed to present the physics of heavy fermions materials\cite{hewson}
(see next section). Our approach will be numerical, through the density-matrix
renormalization group (DMRG) \cite{white} and the Lanczos technique.\cite{dagottorev}
These techniques are non-perturbative, however limited by the system
size. For this reason, we consider the two-leg Kondo ladder (2-LKL),
which is the simplest geometry able to present an orbital order.

The orbital order, also called flux or orbital current phase, has
already been discussed in the context of the high temperature superconductors.
The standard two-leg t-J ladders model present a short-range orbital
order,\cite{rungwhite} while an extended version has long-range orbital
order for some parameters. \cite{obitaltroyer} A recent detailed
discussion of the orbital order in the context of a Hubbard model
can be found in Ref. \onlinecite{currmarston}.

We close this section mentioning that a model very similar to the
KLM was used to describe the magnetism of $URu_{2}Si_{2}$. Sikkema
and collaborators\cite{Sikkema1}, through a mean field calculation,
showed that the Ising-Kondo lattice model with transverse field presents
a weak ordered moment, similar to the one observed in experiments.
However, the Ising-KLM model was not able to reproduce the large specific
heat jump.

\section{MODEL}

In order to investigate the heavy fermion systems the minimum ingredients
that a microscopic model must consider are two types of electrons,
the conduction electrons in the s-, p-, or d-, orbital as well the
electrons in the inner f-orbitals. \cite{Tsunetsugu} In the literature
there are two well known standard models that consider these two kind
of electrons, the periodic Anderson model (PAM) and the KLM.\cite{Tsunetsugu}
In an appropriate parameter regime (mainly (i) the mobility of the
$f$ electrons is very small, which is relevant for the heavy fermion
system and (ii) that the Coulomb interaction of the electrons in the
$f$ orbitals is very large) Schrieffer and Wolff\cite{Tsunetsugu,schriefferwolff}
showed that the KLM can be derived from the PAM. We consider in this
work the KLM which has less degrees of freedom per unit cell than
the PAM and it is easier to explore numerically.

The KLM incorporate an interaction between the localized spins and
the conduction electrons via exchange interaction $J$. To attack
this model in two or three dimension by unbiased non-perturbative
numerical approaches is an impossible task at the present moment.
However, it is possible to consider quasi-one-dimensional systems
such as the N-leg ladders model.

We consider the 2-LKL with $2$x$L$ sites defined by\[
H_{KM}=-\sum_{<i,j>,\sigma}(c_{i,\sigma}^{\dagger}c_{j,\sigma}^{\phantom{\dagger}}+\mathrm{H.}\,\mathrm{c.})+J\sum_{j}\mathbf{S}_{j}\cdot\mathbf{s}_{j}\]

\begin{equation}
+J_{AH}\sum_{<i,j>}\mathbf{S}_{i}\cdot\mathbf{S}_{j},\end{equation}
 where $c_{j\sigma}$ annihilates a conduction electron in site $j$
with spin projection $\sigma$, $\mathbf{S}_{j}$ is a localized spin
$\frac{1}{2}$ operator, $\mathbf{s}_{j}=\frac{1}{2}\sum_{\alpha\beta}c_{j,\alpha}^{\dagger}\bm{\sigma}_{\alpha\beta}c_{j,\beta}^{\phantom{\dagger}}$
is the conduction electron spin density operator and $\bm{\sigma}_{\alpha\beta}$
are Pauli matrices. Here $<ij>$ denote nearest-neighbor sites, $J>0$
(when the KLM is deduced from the PAM obtain $J>0$\cite{Tsunetsugu})
is the Kondo coupling constant between the conduction electrons and
the local moments and the hopping amplitude was set to unity to fix
the energy scale.

We also consider an interaction between the localized spins $J_{AH}$,
we choose $J_{AH}>0$ since antiferromagnetims had been observed in
$URu_{2}Si_{2}$.\cite{Broholmetal} The same model above also represents
the manganites when $J<0$.\cite{livroelbio} In this latter case,
the interaction between the localized spins seems to be important
to stabilize some phases.\cite{livroelbio} This is the motivation
to also consider this interaction. Note that several others terms
in the Hamiltonian could also be included, like the Coulomb interaction
of the electrons in the conduction band, extra electrons hopping,
etc. However, at the present moment, there are no evidences indicating
that such extra terms are relevant to the low energy physics of the
heavy fermion systems. Up to now, it is  well established that
$J$ is essential to describe the magnetism observed in the heavy
fermion systems. At small values of $J$, an antiferromagnetic long-range
order (LRO) is expected due the Ruderman-Kittel-Kasuya-Yosida interaction,
whereas for large $J$ a paramagnetic phase emerges. Doniach\cite{Doniach}
was the first to point out the existence of a quantum critical point
(QCP) due the competition between these two phases.

Unlike other models, such as the $t-J$ model, much less is known
about the Kondo lattice model. Even in the one dimension version,
where the ground state of the Kondo chain is quite well known\cite{Tsunetsugu}
(see also Ref. \onlinecite{gulacsi1dkondo}). New phases have been
reported recently, such as a new ferromagnetic phase\cite{mccullochetal}
inserted into the paramagnetic phase as well as a dimerized phase
at quarter-filling. \cite{dimer} The latter has been questioned recently
by Hotta and Shibata. \cite{commshiba} Those authors claim that the
dimerized phase is an artifact of the open boundary conditions. Indeed,
the boundary condition is very important, as well as the number of sites
considered. In Ref. \onlinecite{commshiba} the authors observed,
mainly, that with an odd number the sites the dimer state does not
exist. The parity of the number of sites is thus very relevant and
an odd number destroys the dimerization.\cite{xavmirpu}

In quasi-one-dimensional systems, such as the N-leg ladders, very
few non-perturbative studies have been reported. Recently, quantum
Monte Carlo\cite{assaad} (QMC) and DMRG\cite{xavierqcp} calculations
of the half-filled Kondo lattice model in small clusters found the
existence of a quantum critical point (QCP) at $J\sim1.45$, in agreement
with previous approximated approaches\cite{kondoqcp,kondoqcp2} (see
also Ref. \onlinecite{qcp_monthoux}). Note that the QMC calculations
were feasible \emph{only} at half-filling, where the famous sign problem
is absent. Moreover, the DMRG results of the $N$-LKL at half-filling
show that the spin and charge gaps are nonzero for \emph{any} number
of legs and Kondo coupling $J$. These results are quite different
from the well known N-leg Heisenberg ladders were the spin gap is
zero for an even number of legs. \cite{dagottosc1}

The phase diagram of the 2-LKL has also been explored numerically.
\cite{2lkphase} In this case, a ferromagnetic phase was observed
only for small densities, very distinctively from the phase diagram
of the 1D Kondo lattice chain, where the ferromagnetism is present
at all electronic densities for large values of $J$. However, it
is similar to the mean field phase diagram of the 3D Kondo lattice
model.\cite{kondopd} In this sense, the 2-LKL presents a better signature
of the phases appearing in real systems than its one-dimensional version.
Interesting that it was also observed dimerization in the 2-LKL\cite{2lkphase}
at conduction electron densities $n=1/4$ and $n=1/2$. As in the
one-dimension version, the RKKY interaction explains these unusual
spin structures. In fact, in some real heavy fermion systems some
unusual spin order structures have indeed been observed. \cite{granadospin}

Here, we consider electronic densities $n$ larger than $0.4$, where
a paramagnetic phase have been observed. \cite{2lkphase} In particular,
we focus on the electronic densities $n=1$ and $n=0.8$. We choose
these densities since the magnitude of the rung-rung current correlation
is bigger for larger electronic densities. We investigate the model
with the DMRG technique under open boundary conditions and use the
finite-size algorithm for sizes up to $2\times L=120$, keeping up
to $m=1600$ states per block in the final sweep. The discarded weight
was typically about $10^{-5}-10^{-7}$ in the final sweep. We also
cross-checked our results with Lanczos technique for small systems.

\section{RESULTS}

Before presenting our results, we briefly discuss the order parameter
associated with a circulating current phase. Such a phase breaks rotational,
translational as well as time reversal symmetries. The appropriated
order parameter to detect this phase is the current between two nearest-neighbour
sites, i. e., $\left\langle \hat{J}_{l,j}\right\rangle $ where the
current operator between two nearest-neighbours $i$ and $j$ is given
by

\[
\hat{J}_{l,j}=i\sum_{\sigma}(c_{l,\sigma}^{\dagger}c_{j,\sigma}-c_{j,\sigma}^{\dagger}c_{l,\sigma}),\]

Strictly speaking, a spontaneous symmetry breaking only appears in
the thermodynamic limit. Only in this limit $\left\langle \hat{J}_{l,j}\right\rangle \ne0$
in the ordered phase. The signature of a spontaneous symmetry breaking
appears in the two point correlation function of the operator that
measures the symmetry. We utilize this fact to infer about the orbital
order. If a continuous symmetry is broken, no long-range order exist
at \emph{finite} temperature in one and two dimensions, as stated by the
Mermin-Wagner-Hohenberg theorem.\cite{Auerbach} At zero temperature
a true long-range order is still possible in two dimensions, while
in one dimension only a quasi-long-range order can manifest, i. e., the
two point correlation function decay algebraic. However, if the translational
symmetry (a discrete symmetry) is broken, even in one dimension a
true long-range order may exist (a famous example is the dimerized
phase of the Majumdar-Ghosh model\cite{majumdarghosh1,majumdarghosh2,majumdarghosh3}
). Since the translational symmetry is broken in the orbital phase,
a true long-range order may occurs in the ground state of the 2-LKL.

In order to observe any trace of orbital order in the ground state
wave function of the 2-LKL, we measure the rung-rung current correlations
defined as

\[
C(l,k)=\left\langle \hat{J}(l)\hat{J}(k)\right\rangle ,\]
 where $\hat{J}(l)$ is the rung current operator for the $l$th rung
given by

\begin{equation}
\hat{J}(l)=i\sum_{\sigma}(c_{l2,\sigma}^{\dagger}c_{l1,\sigma}-c_{l1,\sigma}^{\dagger}c_{l2,\sigma}),\end{equation}
 and $c_{l\lambda,\sigma}$ annihilates a conduction electron on rung
$l$ and leg $\lambda=1,2$ with spins $\sigma=\uparrow\downarrow$.
Since we work with open boundary conditions it is convenient to define
an averaged rung-rung current correlation in order to minimize boundary
effects. We have defined the averaged rung-rung current correlation
as

\begin{equation}
C(l)=\frac{1}{M}\sum_{|i-k|=l}\left\langle \hat{J}(i)\hat{J}(k)\right\rangle ,\end{equation}
 where $M$ is the number of site pairs $(i,k)$ satisfying $l=|i-k|$.
Typically, $M$ in our calculation vary from 3 to 10.

There is a true long-range orbital order if $lim_{l\rightarrow\infty}C(l)\ne0$.
Through this criterion, we can infer about the existence of the orbital
order by measuring $C(l)$ at large distances. If $C(l)$ has an exponential
decay, the linear-log plot shows a linear decay. On the other hand,
if $C(l)$ has an power law decay, the log-log plot present a linear
decay.

We also measure the cosine transform of $C(l)$, i. e.,

\[
N(q)=\sum_{l=1}^{L}C(l)\cos(lq),\]
 in order to infer about periodicity of the oscillatory part of $C(l)$.

In the next two subsections we investigate these correlations for
the 2-LKL at half-filling and close to half-filing, respectively.
We did not find any evidence of long-range orbital order in the ground
state of the 2-LKL. Our results support that the rung-rung current
correlation has an exponential decay at half-filling. Close to half-filling
our results indicate an incommensurate quasi-long range orbital order.

\begin{figure}[!t]
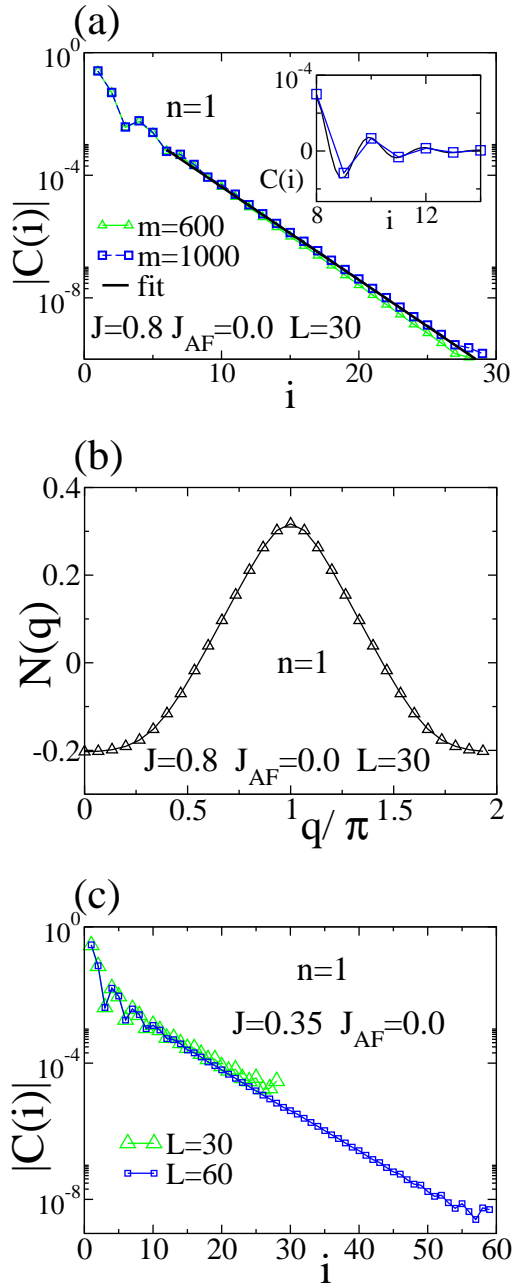

\begin{center}\includegraphics[%
  scale=0.27]{fig2a.eps}\end{center}

\begin{center}\includegraphics[%
  scale=0.27]{fig2b.eps}\end{center}

\begin{center}\includegraphics[%
  scale=0.27]{fig2c.eps}\end{center}

\caption{\label{fig2} (Color online) (a) The linear-log plot of $|C(l)|$
for two distinct value of $m$ with $L=30$ at half-filling. The solid
line in Fig. 2(a) correspond to a fit of Eq. 2 with $\xi=1.43$ and
$a_{0}=0.06$, the RMS per cent error is 0.18. Inset: $C(l)$ vs distance
with $m=1000$. Only few sites are presented. The couplings are $J=0.8$
and $J_{AH}=0$. (b) The cosine transform $N(q)$ of $C(l)$ presented
in Fig. 1(a) with $m=1000$. (c) The linear-log plot of $|C(l)|$
for two distinct size, both with $m=1000$. The couplings are $J=0.35$
and $J_{AH}=0$.}
\end{figure}

\begin{figure}[t]
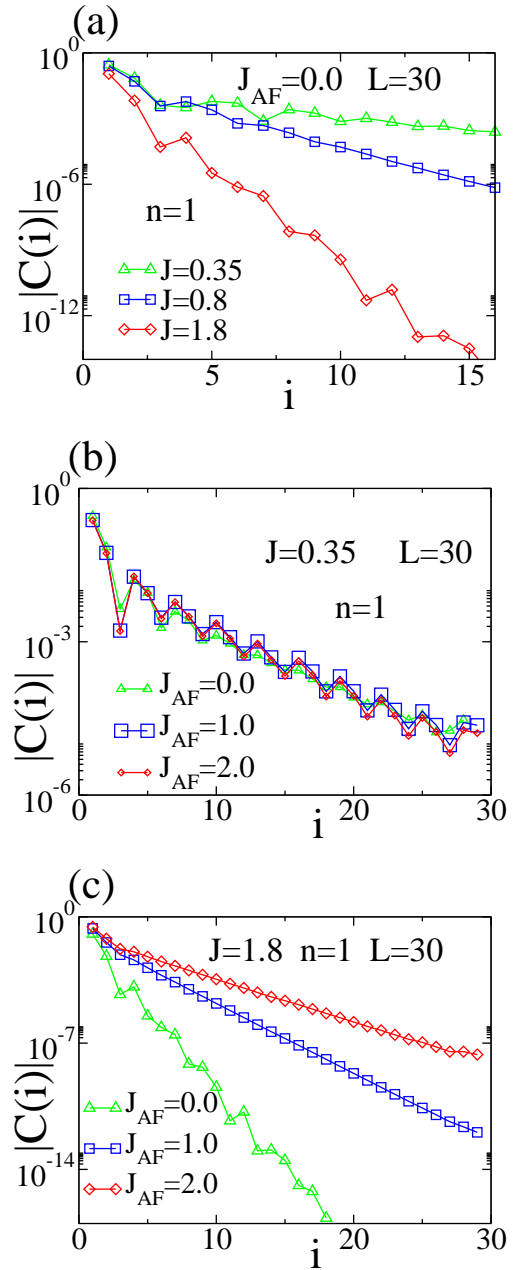

\begin{center}\includegraphics[%
  scale=0.27]{fig3a.eps}\end{center}

\begin{center}\includegraphics[%
  scale=0.27]{fig3b.eps}\end{center}

\begin{center}\includegraphics[%
  scale=0.27]{fig3c.eps}\end{center}

\caption{\label{fig3} (Color online) The linear-log plot of $|C(l)|$ for
a set of representative values of $J$ and $J_{AH}$ for $L=30$ at
half-filling. (a) $|C(l)|$ for $J_{AH}=0$ and $J=0.35,$ 0.8 and
1.8. (b) $|C(l)|$ for $J=0.35$ and some values of $J_{AH}$. (c)
Same as (b) but for $J=1.8$. }
\end{figure}

\subsection{Half-filling }

We start presenting some results for the conduction density $n=1$.
We observed, in this case, that the averaged rung-rung current correlation
behaves as

\begin{equation}
C(l)=a_{0}(-1)^{l}\exp(-l/\xi),\end{equation}
 for all values of $J$ and $J_{AH}$ explored in this work. In Fig.
2(a), we present a typical example of the magnitude of $C(l)$ at
half-filling for a system size $L=30$. As we see, our results indicate
strongly that $C(l)$ has an exponential decay due to the linear decay
in the linear-log plot. The inset in Fig. 2(a) also shows that $C(l)$
is staggered. The solid line in Fig. 2(a) correspond to a fit of Eq.
2 with $a_{0}=0.16$ and a decay length $\xi=1.43$.\cite{comm} We performed a
least-squares fitting, resulting in a root mean square (RMS) of 0.0018
and a correlation coefficient of 0.996. We found that $C(l)$ has a
very small dependence on the number $m$ of states retained in the
truncation process for $J>0.8$, as can be observed in Fig. 2(a).
For $J<0.8$ is very hard to get accurate results, however even for
small $J$ we believe to have captured the correct qualitative behaviour.
Nevertheless, we present most of our results for $J>0.8$, where the
results are more accurate. 

The signature of the sign alternation is observed through the cosine
transform of $C(l)$. In Fig. 2(b), we show the cosine transform of
$C(l)$ present in Fig. 2(a) with $m=1000$. Clearly, we observed
a peak at $q=\pi$ due the sign alternation of $C(l)$. Note that
the finite-size effects are small, as can be seen in Fig. 2(c). For
this reason, we restrict most of our calculus to system size 2x30
in order to save computational time. 

Our results indicate that for small $J$, where the RKKY is expected
to be dominant, the rung-rung current correlations has a bigger correlation
length as we see in Fig. 3(a). On the other hand, for large $J$,
which favors formation of singles, the correlation length is smaller.
This result is expected, since for $J\rightarrow\infty$ the rung-rung
current correlations must go to zero. 

\begin{figure}
\begin{center}\includegraphics[%
  scale=0.29]{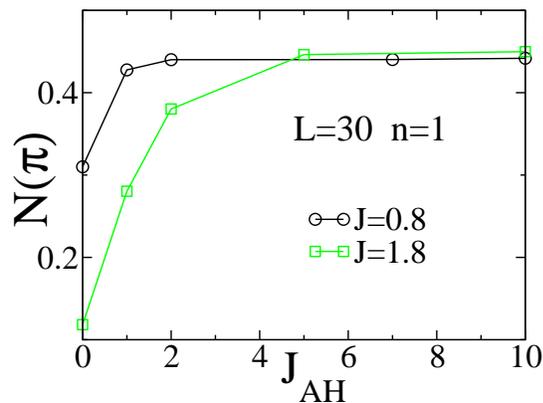}\end{center}

\caption{\label{fig4} (Color online) The peak intensity of $N(q)$ at $q=\pi$
for $J=0.8$ and $J=1.8$ as a function of $J_{AH}$ for a system
size $L=30$ at half-filling.}
\end{figure}

The Hamiltonian Eq. (1) with $J_{AH}=0$ does not lead to a long-range orbital
order at half-filling, as we have observed. Since $J_{AH}$ seems
to be important to stabilise some phases for $J<0$\cite{livroelbio}
, it may be possible that it also stabilises the orbital phase for
$J>0$. For these reason, we also investigate the effect of $J_{AH}$
in the ground state of the 2-LKL. As we see in Fig. 3(b), for small
values of $J$, $J_{AH}$ does not affect significantly $C(l)$. On
the other hand, for larger $J$ as shown in Fig. 3(c), $J_{AH}$ clearly
enhances the length correlation. Although $J_{AH}$ enhanced $C(l)$,
at half-filling only short-range orbital order is observed for several
parameters investigated. 

At half-filling, for all parameters studied, $N(q)$ always presents
a peak at $q=\pi$. In Fig. 4, we present this peak intensity for
$J=0.8$ and $J=1.8$ as function of $J_{AH}$. As we see, the peak
intensity increases with $J_{AH}$ and saturates for large $J_{AH}$
around $\sim0.45$. 

Our main conclusion, for the half-filling case, is absence of long-range
orbital order. Note that it may be possible that the inclusion of
the Coulomb interaction between the electrons in the \emph{conduction
band} leads the system into a phase with long-range orbital order,
as occurs in an extend $t-J$ model. \cite{obitaltroyer}This is under
investigation in the present moment by one of the authors.

\begin{figure}
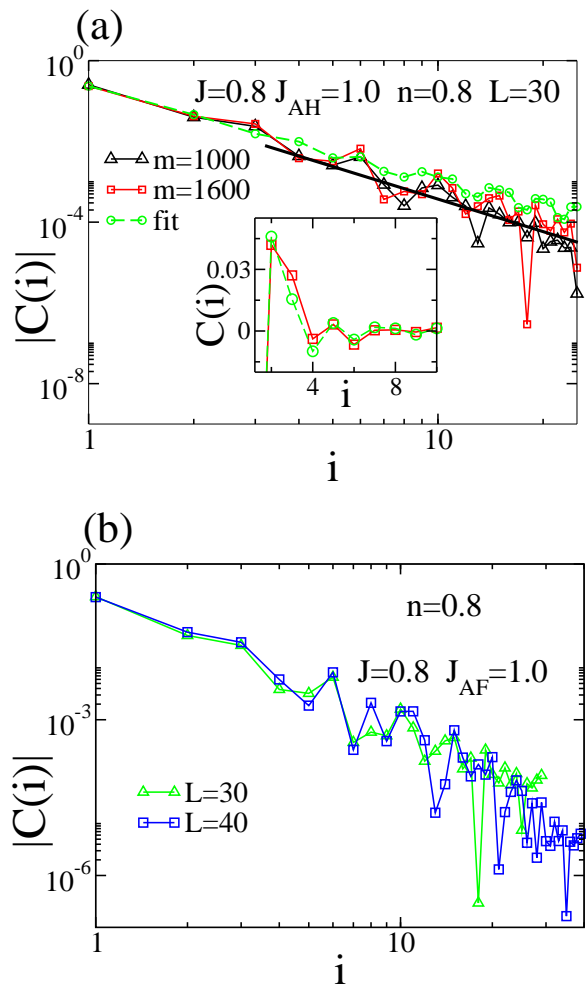

\begin{center}\includegraphics[%
  scale=0.32]{fig5a.eps}\end{center}

\begin{center}\includegraphics[%
  scale=0.32]{fig5b.eps}\end{center}

\caption{\label{fig5} (Color online) The log-log plot of $|C(l)|$ for the
conduction density $n=0.8$. (a) $|C(l)|$ \emph{vs} distance for
$J=0.8$, $J_{AH}=1.0$, and $L=30$ for two distinct value of $m$.
The dashed line is the fit using Eq. (1) with $\alpha_{1}=2$, $\alpha_{2}=3$,
$a_{0}=0.16$ and $a_{1}=-0.33,$ the RMS per cent error is 6 with
a correlation coefficient of 0.998. The solid line is guide by eyes.
(b) $|C(l)|$ for the same parameters of Fig. 5(a) and two distinct
sizes, both cases with $m=1600$. Inset: $C(l)$ vs distance with
$m=1600$. Only few sites are presented.}
\end{figure}

\subsection{Close to half-filling}

Away from half-filling the DMRG calculation of $C(l)$ is less stable,
for this reason we consider system sizes smaller than 2x40 and keeping
up to $m=1600$ states in the truncation process. Although we obtained
results for a few densities away from half-filling, we focus on 
density $n=0.8$ where the magnitude of $C(l)$ is larger. For small
densities is very hard to get accurate results since the current intensity
is very small. In Fig. 5(a), we present the log-log plot of $|C(l)|$
at conduction density $n=0.8$ for a system size $2\times30$ with
$J=0.8$ and $J_{AH}=1.0$ for two different values of $m$. Since
in the log-log plot we obtain a linear decay (see the solid line in
this figure) $C(l)$ must have a power law decay. If we use a linear-log
plot our data does not have a linear decay. As can be seen from that
Figure, it is very hard to get good accuracy even working with $m=1600$
states. Although we were not able to obtain the current-current correlations
at large distances with a high accuracy, we believe to have captured
the correct behavior, {\it i.e.}, a power law decay. The large oscillations
appearing in those Figures are due to fact that some values of $C(l)$
are very close to zero.

Since our data of $C(l)$ in the log-log plot strongly suggest a power
law decay close to half-filling (note that for the half-filling case
the decay is exponential) we tried to fit $C(l)$ with the function 

\begin{equation}
C_{fit}(l)=a_{0}\frac{\cos(n\pi l)}{l^{\alpha_{1}}}+a_{1}\frac{\cos(2n\pi l)}{l^{\alpha_{2}}},\label{eq:1}\end{equation}
 where $n=0.8$ is the density. The dashed curve in Fig. 5(a) corresponds to
a fitting of our data with $m=1600$. We were not able to reproduce precisely
$C(l)$, however the general behavior is quite well described. 

Note also that finite-size effect are larger away from half-filing,
as we can see by comparing the Figs. 5(b) and 2(c). It is important
to mention that we observed, away from half-filling and in very few
distances $l$, that the sign of the averaged correlation C(l=|j-k|)
does not has the same sign of $C(j,k)$ for some pairs of (j,k) satisfying
l=|j-k|. This does not seem to be due to the number of states kept
in the truncation process since we also obtained the same effect for
small clusters with exact diagonalization. 

In Fig. 6 we present $N(q)$ for a representative set of parameters
at conduction density $n=0.8$. As shown in that Figure, there is
no peak at $q=\pi$, signaling an absence of staggered rung-rung current
correlations. For the conduction density $n=0.8$ we observed a cusp
at $q=n\pi$. These results indicate that close to half-filling the
2-LKL presents an \emph{incommensurate} quasi-long-range orbital order. 

\begin{figure}[t]
\begin{center}\includegraphics[%
  scale=0.29]{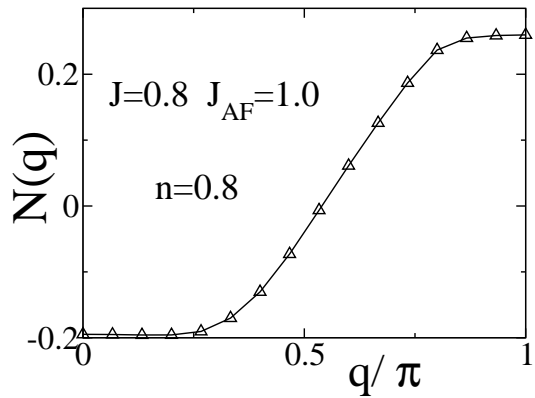}\end{center}

\caption{\label{fig6} The cosine transform $N(q)$ of $C(l)$ presented in
Figure 5(a) with $m=1600$. }
\end{figure}

\section{CONCLUSION}

In this paper, we have investigated the possibility of a two-leg Kondo
ladder present an orbital order. In particular, we focus on
the densities $n=1$ and $n=0.8$. For the several couplings investigated
we did not find any trace of a true long-range orbital order, which would
be relevant to explain the large entropy loss observed in the second order
phase transition of $URu_{2}Si_{2}$ . Our data indicate that the
half-filling case presents a staggered short-range orbital order, while
close to half-filling our results are consistent with an incommensurate
quasi long-range orbital order. Although we did not find evidence
of a long-range orbital order in the ground state of the two-leg Kondo
ladder, we can not yet completely discard this possibility.
It may occur that an extended version of the Kondo lattice model presents
the long-range orbital order. So, we may conclude that either the
orbital phase does not exist and is not the origin of the mysterious
phase transition observed in the the heavy fermion $URu_{2}Si_{2}$
or the standard Kondo lattice Model is not able to reproduce the correct
order observed in the experiments.

\begin{acknowledgments}
The authors thank E. Miranda for useful discussions. This work was
supported by Brazilian agencies FAPESP and CNPq. 

\bibliographystyle{apsrev}
\bibliography{/home/jcandido/FILES/textos/refs_rev4}

\begin{thebibliography}{39}
\bibitem{comm} Since we are interested in the long range behavior of $C(l)$ 
the short distances were disconsidered in the fitting process.
\end{thebibliography}
\end{acknowledgments}

\end{document}